# Study on departure time choice behavior in commute problem with stochastic bottleneck capacity: Experiments and modeling

Dongxu Lu[a], Rui Jiang[a], Ronghui Liu[b], Qiumin Liu[a], Ziyou Gao[a]

[a] *Key Laboratory of Transport Industry of Big Data Application Technologies for Comprehensive Transport, Ministry of Transport, Beijing Jiaotong University, Beijing 100044, China*

[b] *Institute for Transport Studies, University of Leeds, Leeds LS2 9JT, United Kingdom*

**Abstract**：Uncertainty is inevitable in transportation system due to the stochastic change of demand and supply. It is one of the most important factors affecting travelers' choice behavior. Based on the framework of Vickrey's bottleneck model, we designed and conducted laboratory experiment to investigate the effects of stochastic bottleneck capacity on commuter departure time choice behavior. Two different scenarios with different information feedback are investigated. The experimental results show that the relationship between the mean cost ($E(C)$) and the standard deviation of cost ($\sigma$) can all be fitted approximately linearly with a positive slope $\sigma = E(C)/\lambda^* - m$ ($\lambda^* > 0$). This suggests that under the uncertain environment, travelers are likely to minimize their travel cost budget, defined as $E(C) - \lambda^*\sigma$, and $\lambda^* > 0$ indicates that the travelers behave risk preferring. The experiments also found that providing the cost information of all departure times to the commuters lowered the commuters' risk preference coefficient (i.e., $\lambda^*$ decreases). We propose a reinforcement learning model, which is shown to reproduce the main experimental findings well.

## 1. Introduction

Traffic congestion is serious almost in every large city, especially during peak hours. The classic bottleneck model was first proposed by Vickery (1969) to describe people's commuting behavior and traffic congestion evolution in the rush hour. In this model, a fixed number of commuters depart from the same origin (home) to the same destination (work) along a single road. There is a potential bottleneck with a fixed capacity on the road, and it will be active when the departure rate exceeds capacity. As a result, commuters experience queuing and have queuing cost. At the same time, they will be penalized for arriving early or late. Therefore, they face a tradeoff between travel time and schedule delay costs. Commuters adjust the departure time to minimize their trip costs. In the user equilibrium state, no one can unilaterally change the departure time to increase payoff.

The proposal of the Vickery's model laid a solid foundation for characterizing the departure time choice behavior of commuters during the morning peak. Since then, many improved models have been proposed to consider more diverse scenarios, such as pricing (Arnott et al., 1990; Laih, 1994, 2004; Lindsey et al., 2012; Wang and Sun, 2014), elastic demand (Arnott et al., 1993; Yang and Huang, 1997), heterogeneous commuters (Arnott et al., 1994; Lindsey, 2004; van den Berg and Verhoef, 2011; Yao et al., 2012), integration of morning and evening peaks (de Palma and Lindsey, 2002; Zhang et al., 2005; Li et al., 2014), modal split (Tabuchi, 1993; Huang, 2002; Lu et al., 2015), rail transit (Hao et al., 2009; Yang and Tang, 2018), consecutive bottlenecks (Kuwahara, 1990; Lago and Daganzo, 2007), tradable credit scheme (Nie and Yin, 2013; Xiao et al., 2013; Tian et al., 2013), car-pooling (Xiao et al., 2016), ride-sharing (Ma and Zhang, 2017; Wang et al., 2019), automatous vehicles (Liu, 2018; Zhang et al., 2019), time dependent capacity (Zhang et al., 2010), queue dependent capacity (Chen et al., 2019), parking (Tian et al., 2019), and so on.

Commuters face the problem that the supply and demand are not fixed when they make departure choice. Demand sides are mainly from the travel demand fluctuations. Supply sides are due to e.g., traffic accidents, road works, traffic signals and weather, which cause capacity degradation.

In terms of departure time choice with uncertainty, many theoretical works have been reported. For example, Lindsey (1994) considered a general distribution of bottleneck capacity, and studied the properties of no-toll equilibrium and system optimum of the commuting system. Arnott et al. (1999) considered the case where the demand and capacity are both stochastic and examined the effect of information on the total social cost. Xiao et al. (2015) studied the situation that capacity follows uniform distribution. They analyzed four possible departure-time patterns: always early + always queuing, early or late + always queuing, always late + always queuing, always late + possible queuing. Long et al. (2017) generalized the uniform distribution of capacity to general distribution and pointed out that there are two more departure-time patterns: always early + possible queuing, early or late + possible queuing.

In most theoretical works, it was assumed that all travelers minimize their expected cost

$$E(C(t)) = E[\alpha T(t) + \max(\beta(t^* - t - T(t)), 0) + \max(\gamma(t+T(t) - t^*), 0)]. \quad (1)$$

Here $t$ is departure time, $T(t)$ is travel time, $t^*$ is work start time, $C(t)$ is trip cost, $\alpha$, $\beta$ and $\gamma$ denote the unit cost of travel time, the unit cost of schedule delay early (SDE), and the unit cost of schedule delay late (SDL), respectively.

Li et al. (2008, 2009a, 2016, 2017) proposed a different departure time choice principle, in which it is assumed that all travelers minimize

$$u(t) = \alpha E(T(t)) + \max\left(\beta\left(t^* - t - E(T(t))\right), 0\right) + \max(\gamma(t+E(T(t)) - t^*), 0) + \varepsilon\sigma(T(t)). \quad (2)$$

Here $\varepsilon$ is a parameter.

Note that for commuters always arrive early, the expect cost

$$E(C(t)) = (\alpha - \beta)E(T(t)) + \beta(t^* - t). \quad (3)$$

For commuters always arrive late,

$$E(C(t)) = (\alpha + \gamma)E(T(t)) + \gamma(t - t^*). \quad (4)$$

For commuters either early or late, one can derive (Li et al., 2009a, 2016; Fosgerau, 2010; Fosgerau and Karlstrom, 2010)

$$E(C(t)) = \alpha E(T(t)) + \max\left(\beta\left(t^* - t - E(T(t))\right), 0\right) + \max(\gamma(t+E(T(t)) - t^*), 0) + \xi_t \sigma(T(t)), \quad (5)$$

where $0 \leq \xi_t \leq \frac{\beta+\gamma}{2}$ is an attribute-level dependent parameter. Therefore,

$$u(t) = E(C(t)) + \begin{cases} \varepsilon\sigma(T(t)) & \text{if commuters always early or late} \\ (\varepsilon - \xi_t)\sigma(T(t)) & \text{if commuters either early or late} \end{cases}. \quad (6)$$

In other words, Li et al. (2008, 2009a, 2016, 2017) assumed that commuters choose their departure times according to both expected travel cost and the standard deviation of travel time. However, the weight coefficient of the standard deviation of travel time is situation dependent.

Recently, Jiang and Lo (2016) have extensively considered the incentive of a traveler to choose a specific departure time under random travel conditions. They related the influence of travel cost variability on departure time choice and assumed that commuters minimize

$$\bar{u}(t) = E(C(t)) + \lambda \tilde{\sigma}(t), \quad (7)$$

in which $\tilde{\sigma}(t)$ denotes the variability of travel cost and is defined as

$$\tilde{\sigma}(t) = \int_{\theta_{min}}^{\theta_{max}} |C(t) - E(C(t))| f(\theta) d\theta. \quad (8)$$

Here $\theta$ is the random variable, $f(\theta)$ is its probability density function, $\theta_{min}$ and $\theta_{max}$ are lower and upper bound of the random variable, respectively, $\lambda$ is risk attitude parameter.

Finally, we would like to mention that Li et al. (2009b) proposed that a cost function consisting of expected travel cost and variability of travel cost

$$\hat{u}(t) = \chi E(C(t)) + \lambda \sigma(C(t)) \quad (9)$$

can be adopted to model travelers' choice behavior under uncertainty, although they only studied the special case $\lambda = 0$.

To better understand the departure time choice principle of commuters and examine the theoretical assumptions, one can use method of laboratory experiment, which is a powerful tool for studying people's choice behavior and has been widely used to examine equilibrium (Helbing et al., 2002; Gabuthy et al., 2006; Daniel et al., 2009; Rapoport et al., 2014), study traffic paradoxes (Ramadurai and Ukkusuri, 2007; Morgan et al., 2009; Rapoport et al., 2009, 2014) and assess transportation demand management measures (Hartman, 2012; Aziz et al., 2015; Rey et al., 2016).

To our knowledge, there is only one experiment reported on the departure time choice behavior concerning uncertainty. Rapoport et al. (2010) considered the variability of bottleneck capacity when studying batch queue problems. Two ferries with different capacity would arrive on any particular round with equal probability. In their experiment, the subjects chose whether and when to join the queue. In mixed-strategy equilibrium, the player's expected payoff equal to each other. The experimental results show that players' aggregate behavior diverges from mixed-strategy equilibrium in experiment with stochastic capacity. In the uncertain situation, the players are optimistic about obtaining high returns.

Motivated by the fact, this paper performs laboratory experiment to study the departure time choice behavior of commuters in bottleneck model with stochastic capacity. The experimental results show that the relationship between the mean cost and the standard deviation of cost can be fitted approximately linearly with a positive slope. This suggests that under the uncertain environment, travelers are likely to minimize their travel cost budget and they behave risk preferring in the given experiment scenario. Finally, a reinforcement learning model is proposed to simulate the behavior mechanism of the commuters.

The rest of the paper is organized as follows. Section 2 introduces the experimental setup. The experimental results are presented in details in Section 3. In Section 4, a reinforcement learning model is proposed and simulation results are presented. Section 5 gives a conclusion.

## 2. Experiment Design

The experiment was carried out in the computer labs of Beijing Jiaotong University. The interactions were executed via computer and were anonymous. 120 undergraduate students from Beijing Jiaotong University were recruited for the experiments. They were divided into 6 groups and each group has 20 players. The numbers of male and female players are almost equal in each group.

Our experiment is based on a discrete bottleneck model, in which 20 players commute from a single origin (e.g., home) to a single destination (e.g., workplace) along a single road. On the road, there is a potential bottleneck with capacity $s$, which is constant within day but fluctuates from day-to-day following a uniform distribution. If the flow rate exceeds $s$, the bottleneck will be activated and a queue will build up. The players know the distribution of capacity, but do not know the capacity on each specific day.

The travel time from home to workplace is $T(t) = T^f + T^v(t)$, where $T^f$ is the free travel time, $T^v(t)$ is the queuing time due to congestion, and $t$ is the departure time from home. Without loss of generality, we set $T^f = 0$ as usual.

Let $q(t)$ be the queue length. Then, a player's travel time equals the queuing time $T^v(t) = q(t)/s$, in which

$$q(t) = \max(q(t-1) + n(t) - s, 0). \tag{10}$$

Here $q(t-1)$ is the queue length at previous departure time $t-1$, $n(t)$ is the number of players who depart at time $t$.

Given that the working time is $t^*$, according to the bottleneck model, if a player leaves home at time $t$, his/her cost is

$$C(t) = \alpha T^v(t) + \beta \cdot \max(t^* - t - T^v(t), 0) + \gamma \cdot \max(t + T^v(t) - t^*, 0), \tag{11}$$

where the first term on the right-hand side is the queuing cost, the second term is the cost for early arrival, and the third term is the cost for late arrival. The coefficients $\alpha$, $\beta$ and $\gamma$ obey $\gamma > \alpha > \beta$, which is in accordance with Small's empirical results (1982).

We conducted 6 sets of experiments under two different scenarios of feedback information, see Table 1. In the case of personalized information (Scenario A), only information related to that player is provided, which includes information on the departure time the player chose, his early/late arrival time and cost, queuing time and cost, total cost, score in the previous round, and cumulative score in all previous rounds. In the case of general information (Scenario B), costs of all departure times are provided to the player, as well as his/her score in the previous round and the cumulative score. Fig.1 shows the snapshots of experiment screen corresponding to the two scenarios.

The cost is same for the players who choose the same departure time in a particular round, depending on the behaviors of all players. In each round every player was given initial points. Here one round corresponds to one day. At the end of each round, the individual's score was computed by subtracting cost from their initial points.

Table 1 Experimental designs in each scenario

|  | Number of set | Parameter $\alpha/\beta/\gamma/s$ | | Initial points | Information provided |
| --- | --- | --- | --- | --- | --- |
|  |  | $[\alpha\ \beta\ \gamma]$ | $[s_{min}, s_{max}]$ |  |  |
| Scenario A | 3 | [2 1 5] | [1.33, 4.00] | 20 | Personalized |
| Scenario B | 3 | [2 1 5] | [1.33, 4.00] | 20 | General |

In our experiment, the work start time is set to 9 a.m., and there are 16 discrete departure times available for the players to choose from. In each round, each player is asked to select one departure time and then click the 'submit' button. When all 20 players had submitted their choices, we calculate the queuing cost and early/late arrival cost for everyone according to the cost function and the capacity of that round.

At the start of the experiment, we took approximately 15 minutes to explain the game to all players, followed by a Q & A session. Each set runs for 150 rounds and lasts approximately 90 minutes. Although there was no time limitation participants' decision making in each round, it was recommended that participants submit their decisions within 20 seconds (there was a 20 second countdown on the screen). The experiment would only move to the next round if and only if all players have submitted their decisions.

When a set of experiment is over, the score of each player was converted to a payoff (in Chinese Yuan) at a ratio of 100: 3. The payoff plus 20 Yuan show-up bonus was their total income. The mean income of all 6 sets was 72.99 Chinese Yuan.

(a)

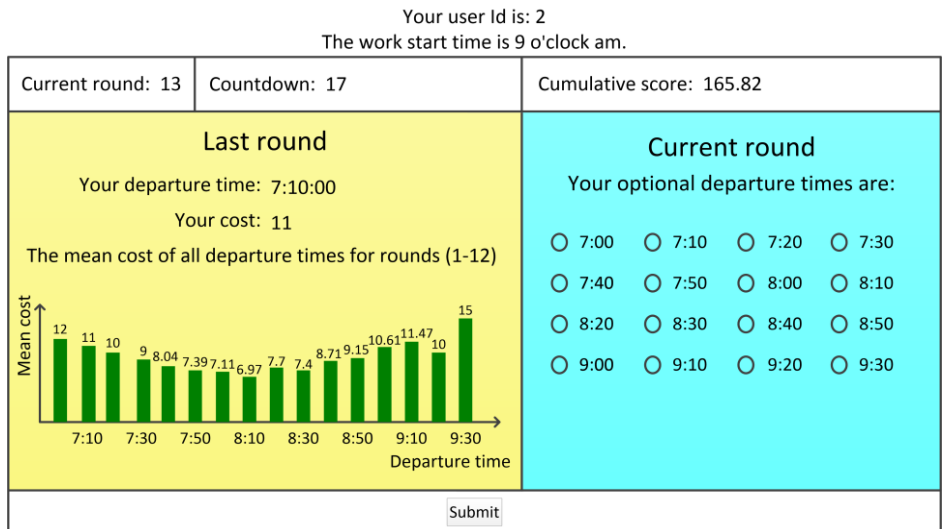

(b)

Fig.1 Snapshot of experiment screen for (a) Scenario A with personalized information; and (b) Scenario B with general information.

## 3. Experimental Result

Take one set of experiment in Scenario A as an example, we first present the number of players and cost at each departure time in each round in Fig.2. It can be seen from Fig.2 (a) that the majority of players choose to depart between 08:00 and 09:00, a small number of players choose to depart at 07:50 and 09:10. Very few depart before 07:50, and no player chooses departure time 07:10, 09:20, and 09:30. The number of commuters in each departure time slot significantly fluctuates throughout the experiment.

The cost at each departure time is presented in Fig.2 (b). It can be seen that with departure times closer to the working start time of 09:00, the cost fluctuation gets higher, and the fluctuation persists until the end.

Summary statistics on the mean number, mean cost and standard deviation of cost, for the 3 sets of experiment in Scenario A, are shown in Fig.3. One can see more clearly that the highest number of players choose to depart at around 8:10, which also corresponds to the lowest mean cost. On the other hand, the standard deviation of cost is the largest around 9:00. The late departure times (at and after 09:00) also correspond to large mean cost, see Fig. 3(b). For early departure times, the cost equals to $\alpha(t^* - t)$, since there is always no queue, irrespective of the capacity. Accordingly, the standard deviation of cost at these early departure times is zero. Similar results are observed in Scenarios B (see Appendix A).

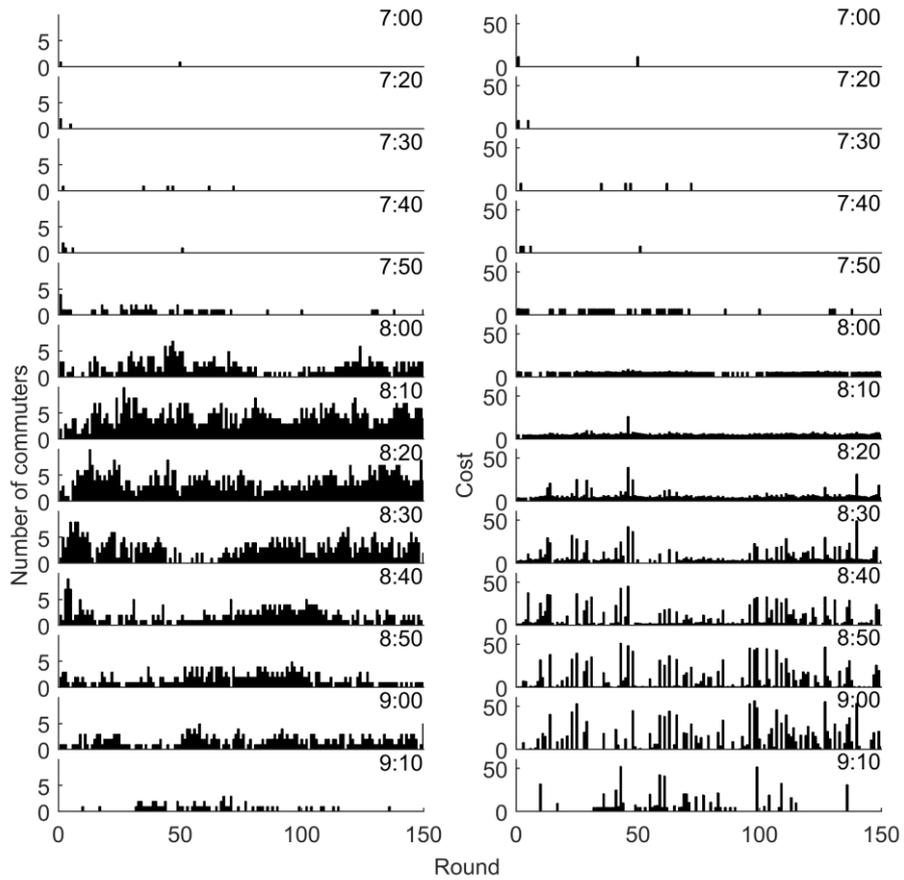

(a) (b)
Fig.2 Number of commuters (a) and cost (b) at each departure time over the 150 rounds in one set of experiment in Scenario A.

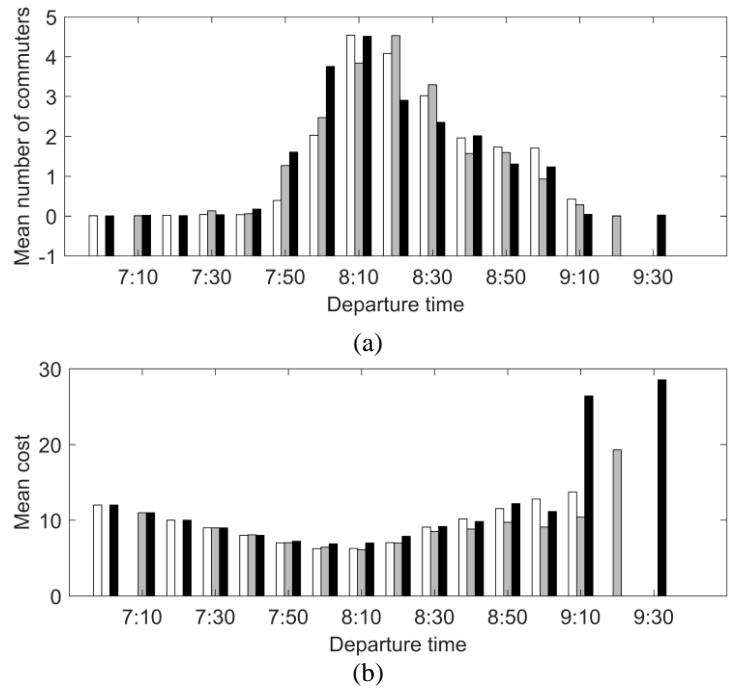

(a)

(b)

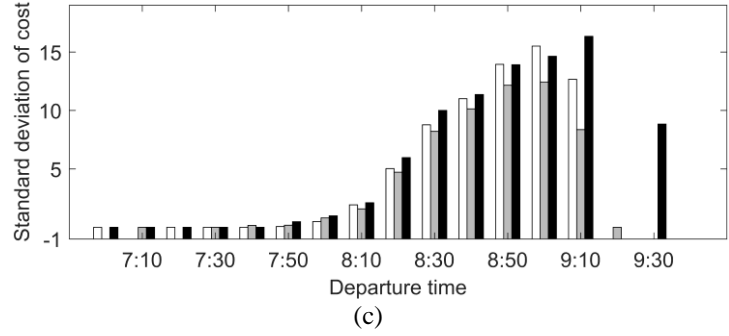

(c)

Fig.3 (a) mean number of commuters, (b) mean cost, and (c) standard deviation of cost in the three sets of experiment in Scenario A. The histograms in the figure represent three sets of experiment in this Scenario. Absence of some bars is because no player chose that departure time.

Fig.4 shows standard deviation of cost vs. mean cost for each departure time in the 6 sets of experiment. The area of each data point is proportional to the number of times that particular departure time was chosen. We make a weighted linear fit of the data, in which weight is set as the area of the data point. One can see that in both Scenarios, the relationship can all be fitted approximately linearly with a positive slope, i.e.

$$\sigma = E(C)/\lambda^* - m, \tag{12}$$

where $E(C)$ is mean cost and $\sigma$ is standard deviation of the cost, and parameters $\lambda^* > 0$ and $m > 0$. A reformulation thus leads to

$$\lambda^* \cdot m = E(C) - \lambda^* \cdot \sigma. \tag{13}$$

Similar to the definition of travel time budget proposed by Lo et al. (2006), we can define $\lambda^* \cdot m$ as travel cost budget (TCB), where $\lambda^*$ is the risk preference coefficient, see also Eq.(9). Our experiment thus demonstrates that faced with uncertain condition (in this case uncertain bottleneck capacity), the players are likely to choose the departure time according to their travel cost budget and they behave risk preferring, with $\lambda^* > 0$.

We also examine the departure time choice principle (2) and (7). Fig.5 shows standard deviation of travel time (i.e. $\sigma(T(t))$) vs.

$$u'(t) = \alpha E(T(t)) + \max\left(\beta\left(t^* - t - E(T(t))\right), 0\right) + \max(\gamma(t + E(T(t)) - t^*), 0)$$

for each departure time in the 6 sets of experiment. Obviously, there is no linear relationship between these $u'(t)$ and $\sigma(T(t))$, indicating that the players in the experiments did not follow this choice principle.

Fig.6 shows $\tilde{\sigma}(t)$ vs. mean cost for each departure time in the 6 sets of experiment. There is also a linear relationship between them, which is very similar to Fig.4, except that the slopes in Fig.6 are smaller than that in Fig.4. This is because both $\tilde{\sigma}(t)$ and $\sigma(C(t))$ reflect variability of travel cost, and $\sigma(C(t))$ is larger than $\tilde{\sigma}(t)$.

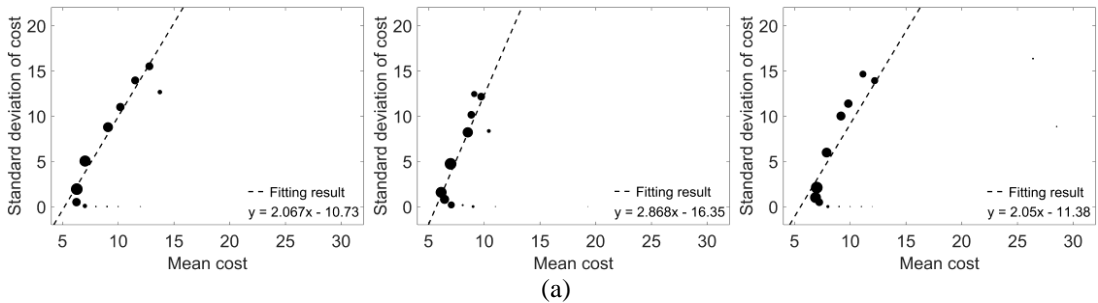

(a)

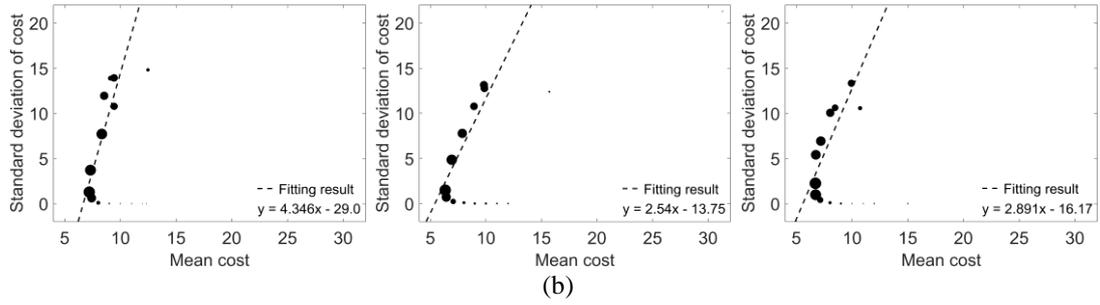

Fig.4 The standard deviation of cost vs. the mean cost in (a) Scenario A, (b) Scenario B. The data points represent the 16 departure times and the size of the data point is in proportion to the number of times that departure time has been chosen.

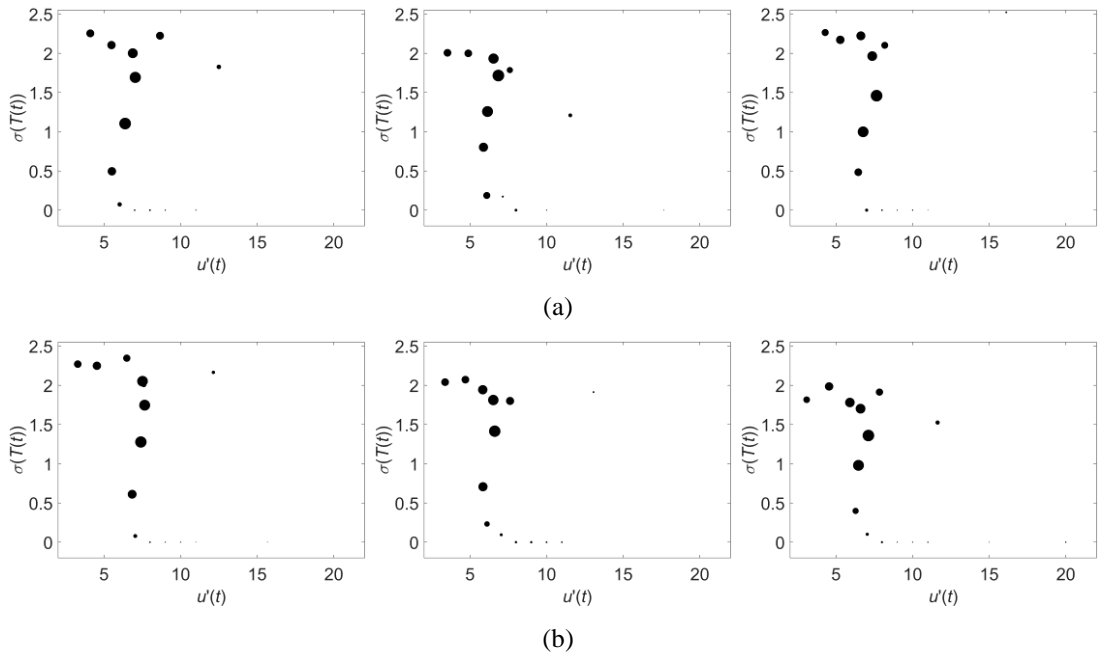

Fig.5 The standard deviation of travel time vs. $u'(t)$ in (a) Scenario A, (b) Scenario B.

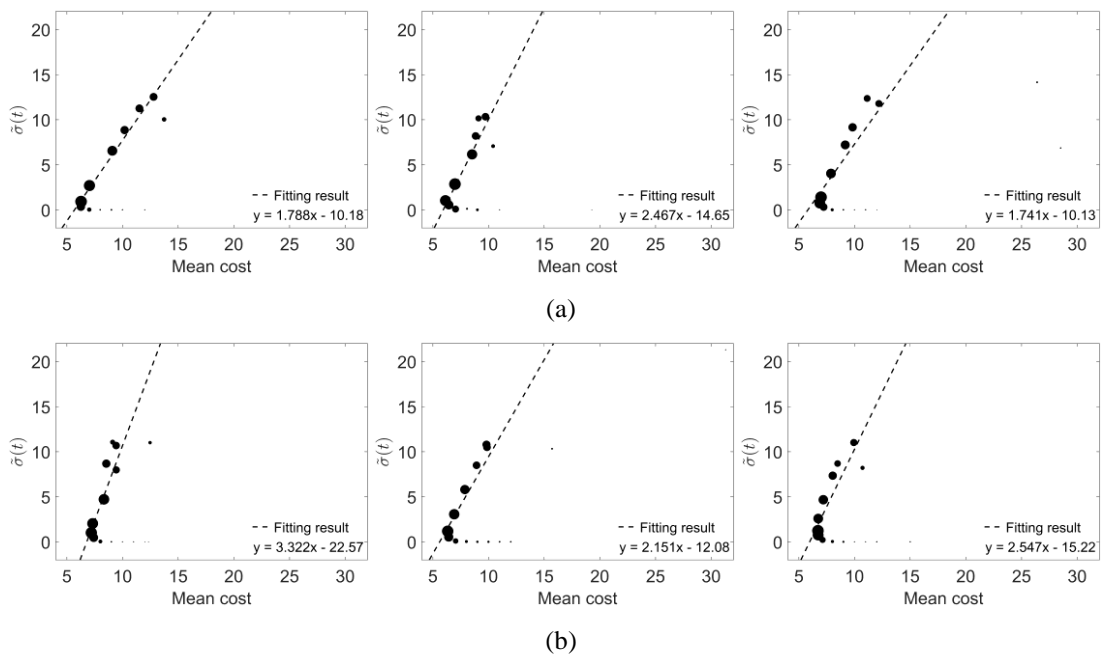

Fig.6 $\tilde{\sigma}(t)$ vs. mean cost in (a) Scenario A, (b) Scenario B.

We investigate the effect of information on the choice behavior. Table 2 compares the mean cost of players and the value of $\lambda^*$ between Scenario A and Scenario B. One can see that providing information of all departure times to the players slightly decreases the commuters' mean cost. At the same time, it decreases commuters' risk preference coefficient (i.e., mean of $\lambda^*$ decreases).

Table 2 A comparison between Scenario A and Scenario B

|  | Scenario A | | | | Scenario B | | | |
| --- | --- | --- | --- | --- | --- | --- | --- | --- |
|  | Set 1 | Set 2 | Set 3 | Average | Set 1 | Set 2 | Set 3 | Average |
| Mean cost of players | 8.769 | 7.782 | 8.721 | 8.424 | 8.377 | 7.784 | 7.843 | 8.001 |
| $\lambda^*$ | 0.455 | 0.333 | 0.477 | 0.422 | 0.281 | 0.364 | 0.358 | 0.334 |

Next we study the dependence of mean cost of all players in round $r$ on the capacity of round $r$. As expected, the mean cost decreases with the increase of capacity, see Fig.7. For the deterministic bottleneck model with fixed capacity, it is known that the cost of each commuter $C'$ depends on capacity $s$

$$C' = \frac{\beta\gamma N}{(\beta+\gamma)s}, \qquad (14)$$

where $N$ is the total number of players. We also plot formulation (14) in Fig.7. One can see that under both Scenarios, when the stochastic capacity is small, the mean cost is significantly larger than $C'$. However, when the stochastic capacity is large, the mean cost is even slightly smaller than $C'$.

The mean cost in Scenario A and Scenario B is 8.424, 8.001, respectively. We calculate the mean cost of the deterministic bottleneck model via

$$\bar{C}' = \frac{1}{s_{max}-s_{min}} \int_{s_{min}}^{s_{max}} \frac{\beta\gamma N}{(\beta+\gamma)s} ds = \frac{\beta\gamma N}{(s_{max}-s_{min})(\beta+\gamma)} \ln \frac{s_{max}}{s_{min}}, \qquad (15)$$

and obtain $\bar{C}'$ is 6.873, much smaller than that in the stochastic capacity situation. This means that if the bottleneck capacity can be predicted and fed back to the commuters in advance, then mean travel cost can be reduced.

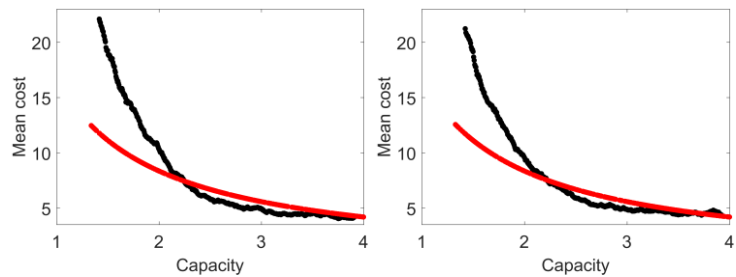

Fig.7 Black line shows mean cost of all players in round $r$ vs. capacity of round $r$ in Scenario A (left) and Scenario B (right). The experimental data are averaged over three sets of experiment and processed by moving average. The red line shows formulation (14).

We examine the choice behavior of each player. Still take one set of experiment in Scenario A as an example, we present the departure time choice of each player in Fig.8 (a) and number the players in ascending order according to their mean cost. One can see that players with low mean cost usually choose early departure times. For example, player 1 has the lowest mean cost. He/she mainly choose 8:00, 8:10, 8:20 and 8:30. Player 2 has the second lowest mean cost. He/she mainly choose 8:10 and 8:30. In contrast, players with high mean cost usually choose late departure times. For example, player 20 has the highest mean cost, he/she mainly choose 8:40, 8:50, 9:00 and 9:10. Player 19 has the second highest mean cost, he/she mainly choose 8:50, 9:00 and 9:10. We examined other 5 sets of experiment, and similar results are observed.

To quantify the relationship, Fig.8 (b) shows the plot of mean cost versus the average departure time. We can see that it has a positive correlation. In addition, in the case of providing general information, the mean cost of the players is more concentrated, indicating that providing information of all departure times has a significant impact on the choice of the players.

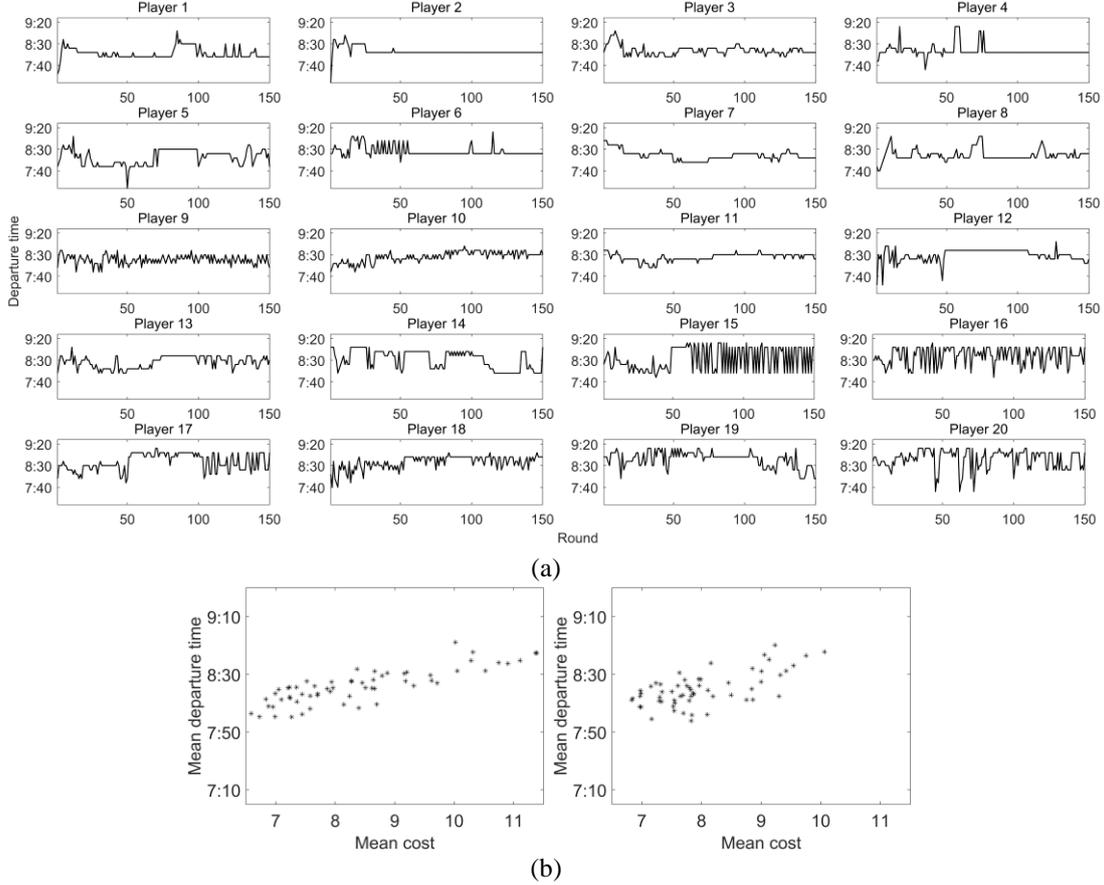

Fig.8 (a) Choice behavior of each subject in one set of experiment in Scenario A. (b) Mean cost vs. the average departure time in three sets of experiment in Scenario A (left), Scenario B (right).

We analyze the impact of capacity in round $r$ on the departure time changing ratio from round $r$ to round $r + 1$. Fig. 9 shows that the departure time changing ratio is negatively correlated with the capacity, and the relationship can also be fitted approximately linearly. Providing information of all departure times increases sensitivity of players (the slope is -0.1623 in Scenario A and -0.1926 in Scenario B).

In Scenario A, the arrival time information has been fed to the players. Next we analyze the relationship between arrival time in round $r$ and the departure time changing behavior from round $r$ to round $r + 1$ in the Scenario. Table 3 shows that after experienced early arrival in last round, most players choose to maintain or delay the departure in next round. Moreover, the closer the arrival time is to 9:00, the smaller the proportion of people who delay the departure. On the other hand, players who were late in the last round tend to depart early. The later the arrival time, the higher the proportion of people who leave early.

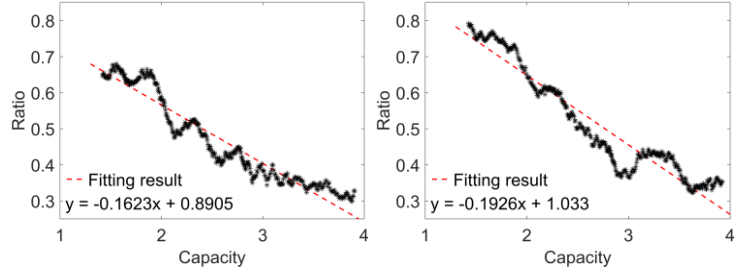

Fig.9 Ratio of departure time changing from round $r$ to round $r+1$ vs. capacity of round $r$ in Scenario A (left) and Scenario B (right). The data are averaged over three sets of experiment and processed by moving average.

Table 3 Departure time changing behavior from round $r$ to round $r+1$ vs. arrival time in round $r$ in Scenario A. The data are averaged over three sets of experiment.

| Arrival time in round $r$ /total number of commuters | Ratio of departing earlier in round $r+1$ | Ratio of not changing departure time in round $r+1$ | Ratio of departing later in round $r+1$ |
| --- | --- | --- | --- |
| Earlier than 7:30 / 13 | 15.38% | 0.00% | 84.62% |
| 7:30 ~ 7:40 / 31 | 6.45% | 51.61% | 41.94% |
| 7:40 ~ 7:50 / 39 | 2.56% | 33.33% | 64.10% |
| 7:50 ~ 8:00 / 441 | 1.36% | 59.18% | 39.46% |
| 8:00 ~ 8:10 / 876 | 2.17% | 58.45% | 39.38% |
| 8:10 ~ 8:20 / 943 | 5.62% | 65.96% | 28.42% |
| 8:20 ~ 8:30 / 1216 | 10.44% | 61.27% | 28.29% |
| 8:30 ~ 8:40 / 1143 | 15.40% | 57.39% | 27.21% |
| 8:40 ~ 8:50 / 1059 | 22.00% | 54.30% | 23.70% |
| 8:50 ~ 9:00 / 943 | 24.50% | 57.58% | 17.92% |
| 9:00 ~ 9:10 / 864 | 42.48% | 46.64% | 10.88% |
| 9:10 ~ 9:20 / 508 | 54.33% | 31.89% | 13.78% |
| 9:20 ~ 9:30 / 309 | 61.49% | 24.27% | 14.24% |
| Later than 9:30 / 555 | 62.52% | 21.44% | 16.04% |

## 4. Reinforcement Learning Model

Our experiment demonstrates that when the capacity is stochastic, commuters are likely to choose the departure time to minimize the travel cost budget instead of the expected cost[1]. Based on this finding, a reinforcement learning (REL) model is proposed to reproduce commuters' choice making behavior under stochastic bottleneck capacity. The REL model has been widely used to model people's choice behavioral characteristics, including route choice (Avineri and Prashker, 2005; Selten et al., 2007; Lu et al., 2014), departure time choice (Daniel et al., 2009; Sun et al., 2017) and travel mode choice (Yang et al., 2017). In the REL model, people have a propensity corresponding to each option of choice, and the propensity is influenced by the choices in previous rounds. Then, the propensity is converted into a probability that controls the choice of people in the next round. In this way, people accumulate experience and learn in the process of making decisions. The REL model emphasizes that people tend to repeatedly choose the strategies which have brought higher payoff in previous rounds.

We propose the REL model as follows

---

[1] Our experiment does not exclude that commuters minimize $\bar{u}(t) = E(C(t)) + \lambda \tilde{\sigma}(t)$. However, since both $\tilde{\sigma}(t)$ and $\sigma(C(t))$ reflect variability of travel cost, and $\sigma(C(t))$ is much more frequently used than $\tilde{\sigma}(t)$, we use travel cost budget in the modeling.

1. Choice in the initial two rounds: it is assumed that in the first and second rounds, all individuals choose each departure time with equal probability.

2. Update propensities: The propensity of individual $i$ in round $r+1$ ($r>1$) is updated by

$$q_i^{r+1}(t) = \begin{cases} E[C_i(t)] - \lambda^* \cdot \sigma[C_i(t)], & t \in V_i \\ \frac{q_i^{r+1}(t_i^k) - q_i^{r+1}(t_i^{k-1})}{t_i^k - t_i^{k-1}}(t - t_i^{k-1}) + q_i^{r+1}(t_i^{k-1}), & t \notin V_i \text{ and } t_i^{k-1} < t < t_i^k \\ \frac{q_i^{r+1}(t_i^2) - q_i^{r+1}(t_i^1)}{t_i^2 - t_i^1}(t - t_i^1) + q_i^{r+1}(t_i^1), & t = t_i^1 - 1 \\ \frac{q_i^{r+1}(t_i^n) - q_i^{r+1}(t_i^{n-1})}{t_i^n - t_i^{n-1}}(t - t_i^n) + q_i^{r+1}(t_i^n), & t = t_i^n + 1 \\ q_i^{r+1}(t_i^1 - 1), & t < t_i^1 - 1 \\ q_i^{r+1}(t_i^n + 1), & t > t_i^n + 1 \end{cases} \quad (16)$$

where $q_i^{r+1}(t)$ is the propensity of individual $i$ to choose departure time $t$ in the $(r+1)^{th}$ round; $V_i = [t_i^1, t_i^2, \cdots, t_i^k, \cdots, t_i^{n-1}, t_i^n]$ is the set of departure times that have been chosen by individual $i$ from round 1 to round $r$, and $t_i^1$ and $t_i^n$ are the earliest and the latest departure time that have been chosen; $E[C_i(t)]$ and $\sigma[C_i(t)]$ are the mean cost and standard deviation of cost at departure time $t$, calculated from the rounds that individual $i$ has chosen departure time $t$; $\lambda^*$ is the parameter representing risk preference.

The basic idea is, for a departure time that has been chosen before, the propensity of individual $i$ is set to equal his/her experienced TCB at that departure time; for departure time between $t_i^1 - 1$ and $t_i^n + 1$ that has not been chosen before, the propensity is obtained from linear interpolation or extrapolation from the propensities of the two nearest chosen departure times; for unchosen departure times earlier than $t_i^1 - 1$ (or later than $t_i^n + 1$), the propensity is set to equal to that on $t_i^1 - 1$ (or $t_i^n + 1$).

3. Update probabilities: The probability of choosing departure time $t$ in round $r+1$ is calculated by

$$p_i^{r+1}(t) = \frac{\exp\left(-\frac{\theta}{\varphi_i} \cdot q_i^{r+1}(t)\right)}{\sum_{k=1}^{T} \exp\left(-\frac{\theta}{\varphi_i} \cdot q_i^{r+1}(k)\right)} \quad (16)$$

where $\theta > 0$ is reinforcement coefficient determining the reinforcement sensitivity, $\varphi_i$ is mean cost of individual $i$ from round 1 to round $r$, $T$ is the number of departure times.

As in experiment, each simulation run includes 20 participants and lasts 150 rounds, using the same capacity sequence as in the experiment. The value of $\lambda^*$ is set to equal to the average value of $\lambda^*$ obtained from experiments ($\lambda^*$ =0.422, 0.334, respectively, in Scenarios A and B). The reinforcement coefficient $\theta$ is the only parameter that needs to be calibrated. The calibration result is $\theta$ =13.9, by minimizing sum of the difference of commuter number at each departure time between experiment and simulation results.

Figs. 10 and 11 compare the experimental and simulation results for the mean number of commuters at each departure time and the average cost in each round in Scenario A. Results in Scenario B are shown in Appendix B. Table 4 shows R-square between simulation and experimental results. One can see that the model can reproduce the collective choice behavior well. Moreover, the REL model also reproduces the linear relationship between standard deviation of cost and the mean cost, as shown in Fig.12.

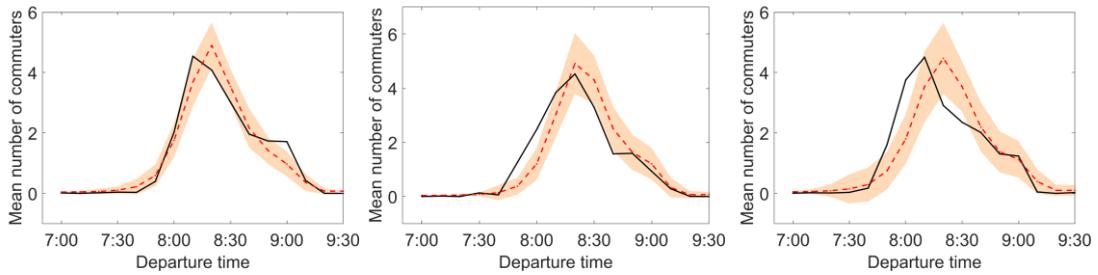

Fig.10 Mean number of commuters at each departure time in experiment and simulation in Scenario A. The black curve is the experimental result, the red dash line is the result of REL model, and the shadow region is the error bar of simulation result.

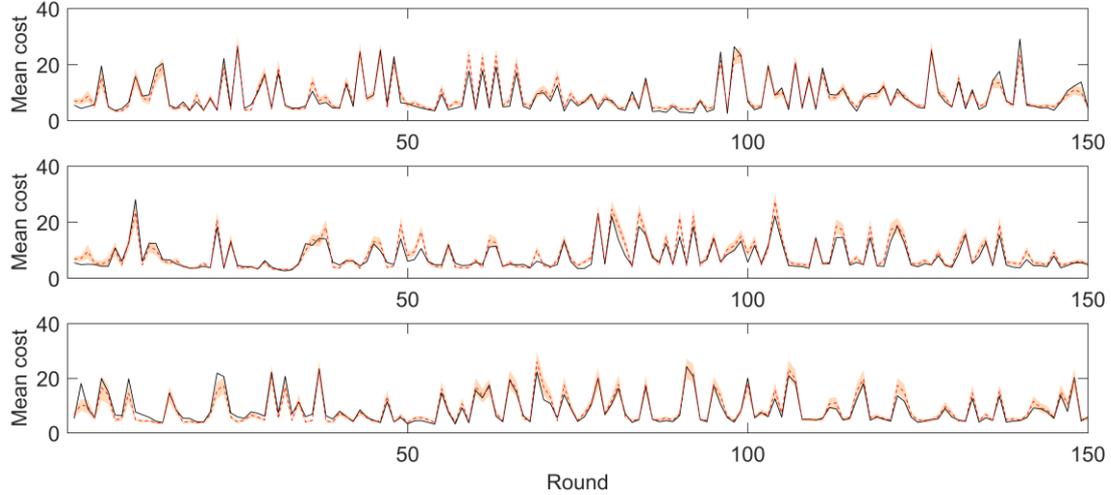

Fig.11 Average cost in each round in experiment and simulation in Scenario A. The black curve is the experimental result, the red dash line is the result of REL model, and the shadow region is the error bar of simulation result.

Table 4 R-square between simulation and experimental results.

|  | Mean number of commuters | | | Average cost | | |
| --- | --- | --- | --- | --- | --- | --- |
|  | Set 1 | Set 2 | Set 3 | Set 1 | Set 2 | Set 3 |
| Scenario A | 0.9290 | 0.8509 | 0.7129 | 0.9309 | 0.8398 | 0.9039 |
| Scenario B | 0.6126 | 0.8872 | 0.7803 | 0.9232 | 0.9066 | 0.8994 |

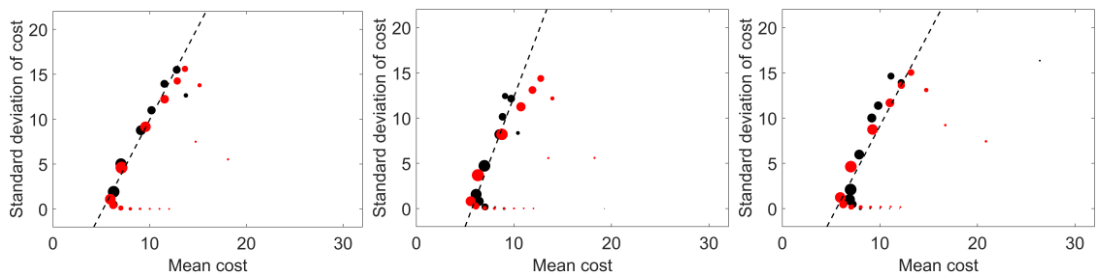

(a)

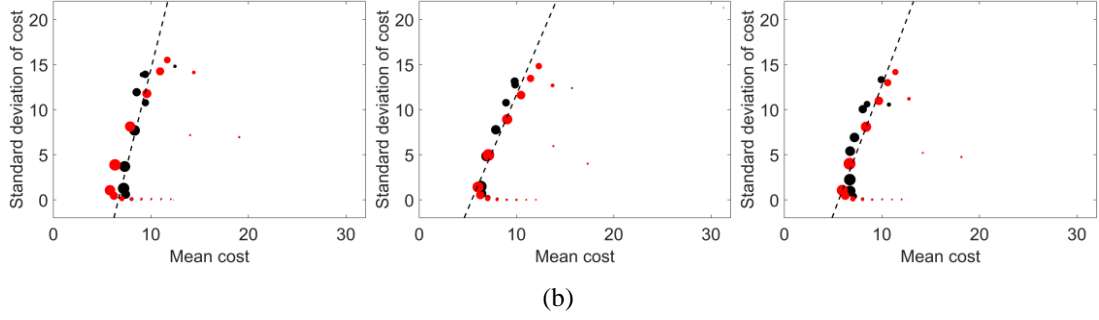

(b)

Fig.12 Standard deviation of cost vs. mean cost at each departure time in experiment and simulation. The area of each point is proportional to the mean number of commuters that choose the departure time. The red points are the simulation results while the black points represent experimental results. The dash line is the fitted curve of the experimental data. (a) and (b) are results of three sets of experiment in Scenario A, B, respectively.

## 5. Conclusion

Uncertainty is an important feature of traffic system, and it is one of the most important factors affecting commuters' choice behavior. Although theoretical works concerning bottleneck model with stochastic capacity have been reported, the departure time choice behavior has not been validated before.

Laboratory experiment provides us with an efficient way to examine behavior characteristic. This method is controllable, repeatable, and low in experimental cost, and thus becomes the preferred method to reveal the potential mechanism of decision-making behavior.

This paper designed and conducted laboratory experiment to investigate the effects of stochastic bottleneck capacity on commuter departure time choice behavior. The most important finding is that relationship between standard deviation of cost and mean cost can be fitted approximately linearly with a positive slope. This suggests that under the uncertain environment, travelers are likely to minimize their travel cost budget, and the travelers behave risk preferring. We would like to mention the overall risk preferring behaviors of players is related to the experimental scenario. It might change if parameters, say $\gamma$, increases.

Other findings include: (i) Providing information of all departure times to the players slightly decreases the commuters' cost. At the same time, it decreases commuters' risk preference coefficient. (ii) In the stochastic capacity situation, the mean cost would increase. (iii) Players with low (high) mean cost usually choose early (late) departure times. (iv) The relationship between departure time changing ratio from round $r$ to round $r + 1$ and capacity in round $r$ can also be fitted approximately linearly with a negative slope. Providing information of all departure times increases sensitivity of players.

We have proposed a reinforcement learning model. The main characteristic of the model is that for some departure times that have not been chosen before, the propensity is obtained from linear interpolation or extrapolation from the propensities of the two nearest chosen departure times. Simulation shows that the model can reproduce the main experimental findings.

The experimental findings would inspire us to propose a TCB based user equilibrium for bottleneck model with stochastic capacity. Actually, we have studied such a problem (Liu et al., 2019), in which the capacity of the bottleneck is constant within a day but changes stochastically from day-to-day between a designed value (good condition) and a degraded one (bad condition). The study revealed that considering variability of travel cost significantly affect departure time choice, and thus needs to be carefully investigated in uncertainty related theoretical and empirical studies.

## Acknowledgment

This work is supported by National Key R&D Program of China (No. 2018YFB1600900), the National Natural Science Foundation of China (Grants No. 71621001, 71631002, 71931002), and Beijing

Natural Science Foundation (Grant No. 9172013).

**Appendix A：Experimental Results in Scenario B**

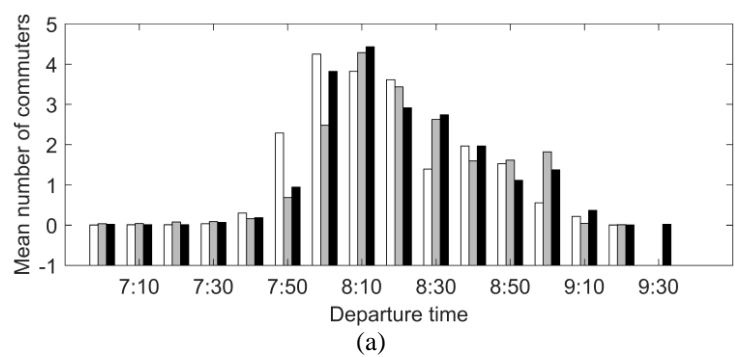

(a)

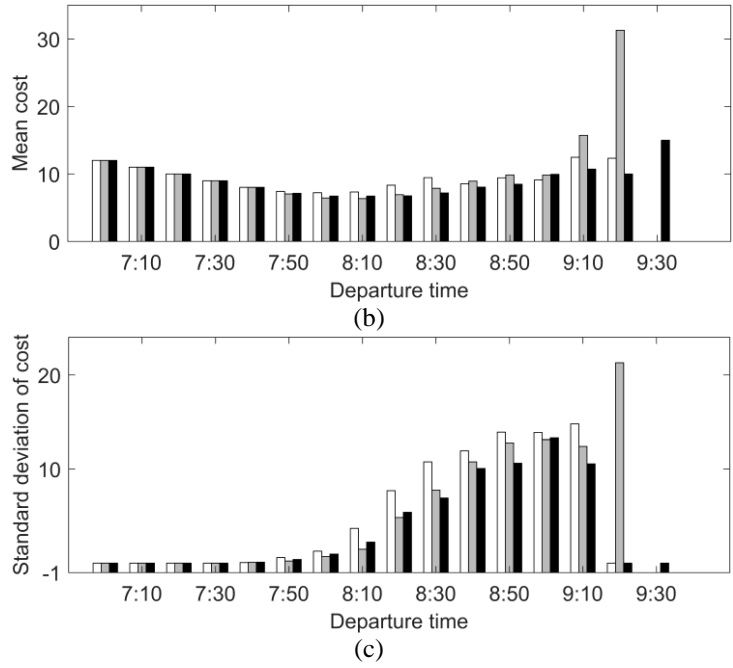

Fig.A1 (a) mean number of commuters, (b) mean cost, and (c) standard deviation of cost in the three sets of experiment in Scenario B. The histograms in the figure represent three sets of experiment in this Scenario.

We would like to mention that in the second set of experiment in Scenario B, one player has misunderstood the cost shown in the interface as payoff. Therefore, he mostly chose very early or very late departure time. Therefore, in the data analysis, the data of this player were removed.

**Appendix B Simulation results in Scenario B**

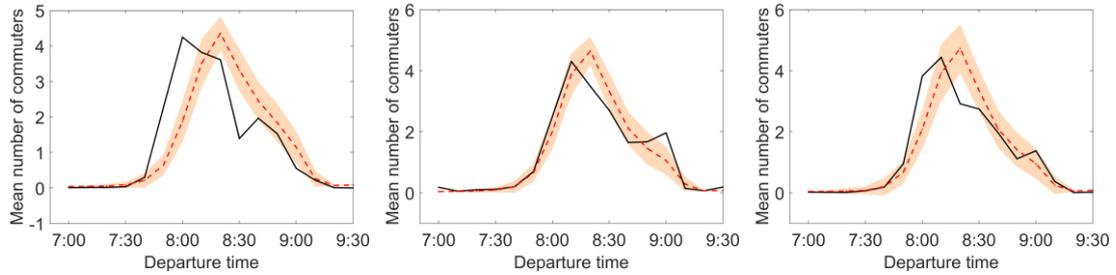

Fig.B1 Mean number of commuters at each departure time in experiment and simulation in Scenario B. The black curve is the experimental result, the red dash line is the result of REL model, and the shadow region is the error bar of simulation result.

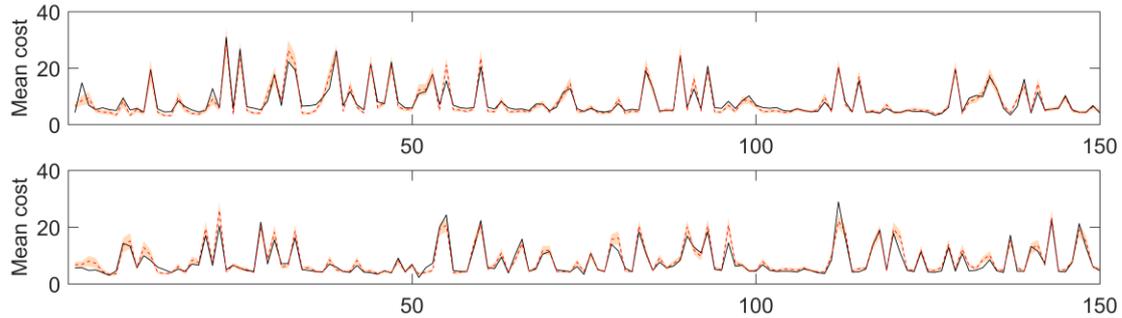

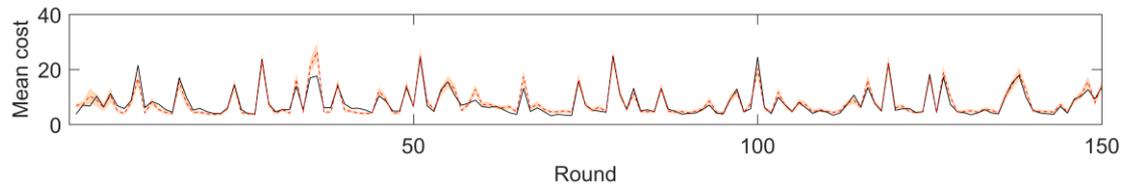

Fig.B2 Average cost in each round in experiment and simulation in Scenario B. The black curve is the experimental result, the red dash line is the result of REL model, and the shadow region is the error bar of simulation result.